\documentclass[12pt]{article}
\newcommand{\ce}[1]{\centerline{#1}}

\begin{document}
\renewcommand{\thefootnote}{\fnsymbol{footnote}}
\vspace*{.5cm}
\begin{center}
{\Large\bf Effective Quark Models in QCD\\ 
at low and intermediate energies}\footnote{Talk given at
11 International Seminar on HEP, QUARKS-2000 (May 2000, Pushkin, Russia) and
at Brittany Workshop on Theoretical HEP (Sept.2000, Guidel, France) }
\end{center}

\vspace{1.0cm}

\ce{\large A.~A.~Andrianov$^{1,2}$, 
V.~A.~Andrianov$^1$, D.~Espriu$^2$ and R.~Tarrach$^2$}

\medskip

\ce{$^1${\it Department of Theoretical Physics,
St.-Petersburg State University,}}
\ce{\it 198904 St.-Petersburg, Russia}
\ce{$^2${\it Departament d'ECM, Universitat de Barcelona}}
\ce{\it 08028 Barcelona, Spain}

\vspace{2.0cm}
\ce{\large\bf Abstract}

\bigskip

\noindent
The effective quark models are employed  to describe the hadronization
of QCD 
in the quark sector. 
They reveal a different structure
depending on how the spontaneous chiral symmetry breaking (CSB) is
implemented.
When the generation of
light pseudoscalar mesons is  manifestly incorporated one deals with
 an extension of the  chiral quark model (CQM)
with the  non-linear realization of chiral symmetry.
If a model is built at the CSB scale by means of perturbation theory 
it generalizes the Nambu-Jona-Lasinio (NJL)
one with chiral symmetry broken due to attractive 4-fermion
forces in the scalar channel.
The matching to high-energy QCD is 
realized at CSB scale by means of Chiral Sum Rules.
Two types of models are compared in their fitting of meson physics.
In particular, if the lowest scalar meson is sufficiently heavy
approaching
the mass of heavy $\pi'(1300)$  then QCD favours an
effective theory which
is dominated by the simplest CQM. 
On the contrary,
a light scalar quarkonium ($m_\sigma \simeq 500MeV$)
supports the NJL mechanism. 

\newpage

\section{Introduction}

The notion of quark model appears in the literature for many different
truncations of QCD: when one makes a request to SLAC-SPIRES service to
FIND TITLE QUARK MODEL the result exceeds 3700 documents. If one assumes
that for each version there are around 100 papers with its application, 
one ends up
with about 30-40 different quark models. Among them there are many which
simulate a part of gluon interaction in the form of a potential or specific
nonlocal forces and thereby do not give a controllable 
truncation of QCD. We leave aside the bulk of those quark models and we
 focus only on
two type of models  which open the way of a systematic 
construction of the low-energy
effective action of QCD in the quark sector. In the models we are considering
 the gluon interaction
is hidden in the effective coupling constants. In 
their simplest form they have been used for a long time 
to reproduce main features of
QCD hadronization [1-4].

The first one is the Chiral Quark Model (CQM) \cite{1}  with
quark fields 
$q_i^\alpha, \bar q_i^\alpha;$ carrying a color, $ i = 1,\ldots, N_c,$
and a flavour, $  \alpha=1,2,..,N_F$ and 
interacting with the colorless chiral field  $U(x)$,
\begin{equation}
{\cal L}_{CQM} =
\bar q \left( i\!\not\!\partial
 -  M_0 (U P_L + U^+ P_R)\right) q
+ \frac14 f^2_0 \mbox{\rm tr}[\partial_{\mu}U\partial^{\mu}U^+],\label{CQM}
\end{equation}
where   $ \not\!\!\partial \equiv \gamma^{\mu}
\partial_{\mu}$ and the projectors $P_{L(R)} = (1 \pm  \gamma_5)/2$ are
used.
$M_0$ is a chirally invariant constituent mass, which has
a non-zero value once chiral symmetry is spontaneously broken. For two light
flavors
$U(x) = \exp \left(i\pi(x)/ F_0\right)$  is a $SU(2)$ matrix
with generators $ \pi \equiv \pi^a T^a$, $a=1,2,3$; corresponding to the 
massless Goldstone bosons -- pions.

The constant $F_0$ is the pion  decay constant, whereas
$f_0$ represents a bare pion decay
constant \cite{2}, which contains
residual gluon contributions not accounted for by the radiative quark effects.

The second approach is provided by  the Nambu-Jona-Lasinio (NJL) model
\cite{3}. It includes a
chirally invariant four-fermion interaction. For massless quarks it reads
\begin{equation}
{\cal L}_{NJL} \, = \,\bar q i\! \not\!\partial  q \,
+\,\frac{g_0}{4N_{c} \Lambda^2}\,
\left[(\bar q q )^{2} - (\bar q \gamma_5 \tau^a q )^{2}\right].
\label{NJL}
\end{equation}
In this model
dynamical breaking of chiral symmetry occurs 
for strong enough coupling  $g_0$. This
leads to the creation of a massless pion state, to the appearance
of a dynamical
quark mass $\Sigma_0$ (similar to $M_0$), and to the generation of a
scalar meson  with the mass $m_{\sigma} \simeq 2\Sigma_0$, the 
$\sigma$-particle.

Both models provide 
the similar set of low energy structural constants describing interactions
between pions and therefore may 
 serve for interpolating
the true low-energy effective action of QCD.

\section{Extension of Quark Models}

For intermediate energies the above
models need to be extended [5-8] in order
to get a
better agreement with meson phenomenology
 and to satisfy the
requirements of QCD.

First, one may build
the most general action describing
strong interactions once the spontaneous CSB
 has taken place at a scale $\Lambda$.
This effective action
should contain the  lightest degrees of freedom
appearing below $\Lambda$, namely the (pseudo) Goldstone
bosons -- pions (for two flavours), assembled in the chiral field $U$ .  
One has to include all local operators -gauge and chiral invariant, 
which can be composed 
of the chiral field $U$ and quarks. Their list contain 
already about 50 operators (see \cite{11}) 
of dim$\leq 6$ but only 
a few of them  contribute to the leading order in the large-cutoff
expansion. 

We embed the  external
sources into the QCD quark lagrangian
in order to compute the correlators of corresponding quark currents,
\begin{equation}
 \hat D \equiv i  \gamma_\mu(\partial_\mu+ \bar V_\mu  + \gamma_5\bar A_\mu) +
 i (\bar S + i \gamma_5 \bar P), 
\end{equation}
where $\langle S\rangle = m_q$, the  matrix of current
quark masses.

To describe the Extended CQM (ECQM) action it is convenient 
to introduce the `constituent' quark
fields $ Q_L \equiv \xi q_L$, $ Q_R \equiv \xi^\dagger q_R$,
$\xi^2\equiv U$,
which transform nonlinearly under  $ SU_L(2) \bigotimes SU_R(2) $
but identically for left and right quark components.

Changing to this basis is given by the following replacements
in the external vector, axial, scalar and pseudoscalar sources
\begin{eqnarray}
&\bar{V}_\mu&  \to  v_\mu = \frac12 \left( \xi^\dagger \partial_\mu \xi -
\partial_\mu \xi  \xi^\dagger +  \xi^\dagger \bar V_\mu \xi +
\xi \bar V_\mu\xi^\dagger - \xi^\dagger \bar A_\mu \xi +
\xi \bar A_\mu \xi^\dagger\right),\nonumber\\
&\bar{A}_\mu&  \to  a_\mu = \frac12
\left( - \xi^\dagger \partial_\mu \xi -
\partial_\mu \xi \xi^\dagger -  \xi^\dagger \bar V_\mu \xi+
\xi \bar V_\mu\xi^\dagger + \xi^\dagger \bar A_\mu \xi +
\xi \bar A_\mu \xi^\dagger\right),\nonumber\\
&\bar{\cal M} &\equiv (\bar S + i 
\bar P)  \to  {\cal M} = \xi^\dagger \bar{\cal M} \xi^\dagger .
\end{eqnarray}
In these variables the relevant part of ECQM action can be represented as, 
\begin{equation}
{\cal L}_{ECQM} = {\cal L}_{ch} + {\cal L}_{\cal M},
\end{equation}
where ${\cal L}_{ch}$ accumulates the interaction of chiral fields and
quarks in the chiral limit in the presence of vector and  axial-vector
external fields, 
\begin{eqnarray} 
{\cal L}_{ch}&=& 
i\bar Q \left( \not\!\! D
 +  M_0 \right) Q
 -\frac{f_0^2}{4}{\rm tr}(a_\mu^2)
+\,\frac{G_{S0}}{4N_{c} \Lambda^2}\,
(\bar{Q}_L Q_R  +
\bar{Q}_R Q_L)^2\nonumber\\  
&&- \frac{G_{P1}}{4N_{c} \Lambda^2}\,
( -  \bar{Q}_L \vec\tau Q_R
+  \bar{Q}_R  \vec\tau Q_L)^2\nonumber\\
 && - \frac{G_{V1}}{4N_c \Lambda^2} \bar Q \vec\tau \gamma_\mu
Q \bar Q \vec\tau \gamma_\mu Q
-
\frac{G_{A1}}{4N_c \Lambda^2} \bar Q \vec\tau \gamma_5\gamma_\mu Q
\bar Q \vec\tau \gamma_5\gamma_\mu Q \nonumber\\
&&+ c_{10} \mbox{\rm tr}[U \bar L_{\mu\nu} U^\dagger \bar R_{\mu\nu}]
, \label{lchir}
\end{eqnarray}
where
\begin{equation}
 Q \equiv Q_L + Q_R,\qquad
\not\!\! D \equiv \not\!\partial  + \not\! v - \gamma_5 \tilde g_A 
\not\! a , 
\end{equation}
with  the `bare' 
axial coupling $\tilde g_A 
\equiv 1 - \delta g_A$ and the `bare' chiral coupling $c_{10}$. 

${\cal L}_{\cal M}$ extends the description 
 for external scalar and pseudoscalar fields and, in particular,
for massive quarks,
\begin{eqnarray}
 {\cal L}_{\cal M} &=&
i (\frac12 + \epsilon) \left(\bar Q_R {\cal M}
 Q_L + \bar Q_L  {\cal M}^\dagger  Q_R \right)
 + i (\frac12 - \epsilon) \left( \bar Q_R {\cal M}^\dagger  Q_L
+  \bar Q_L  {\cal M} Q_R\right)\nonumber\\
&& +  \mbox{\rm tr}\left[  c_0\left({\cal M}  + {\cal M}^\dagger\right) 
 + c_5 ({\cal M} +{\cal M}^\dagger)a_\mu^2 
+ c_8 \left({\cal M}^2 + \left({\cal M}^\dagger\right)^2\right)
\right]  ,
\end{eqnarray}
where the chiral couplings $c_0,c_5,c_8$ (as well as $c_{10}$) 
 are  `bare', different from those introduced in \cite{13}. 
All `bare' coupling constants 
are renormalized by radiative quark contributions and
their physical values  are controlled by the CSR rules (see below).

Thus the effective action suitable for derivation of two-point 
correlators contains 13 parameters to be determined by matching to QCD:
$M_0, \Lambda$(cutoff), bare chiral constants $ f_0, c_0, c_5, c_8, c_{10}$,
the axial pion-quark coupling $\tilde g_A$, the mass asymmetry
$\epsilon$, the four-fermion coupling constants 
$G_{S0}\not= G_{P1}, G_{V1}\not= G_{A1}$. In this approach the pion 
plays a very crucial role, quite different from other hadronic resonances.\\

On the other hand one could construct the QCD effective action
employing the perturbative expansion in the QCD coupling constant $\alpha_s$
for the high-energy gluons and quarks \cite{11b}. 
The effective action for low-energy
quarks is further prepared by means of the derivative (soft-momentum)
expansion, generating an infinite series of quasilocal higher-dimensional
operators.
Then for sufficiently strong couplings,
the new operators may promote the generation of additional
scalar
and pseudoscalar states. These models give an extension of the
linear $\sigma$ model provided by the NJL model,  with the pion being
a broken symmetry partner of the
lightest scalar meson, and with excited pions and scalar
particles, as well as with vector and axial-vector mesons, coming in pairs. 
In particular, when scalar, pseudoscalar, vector and axial-vector
color-singlet channels are examined and dynamical quark masses are 
supposed to
be sufficiently smaller than the CSB cutoff one may 
derive the minimal two-channel
lagrangian in the separable form:
\begin{eqnarray}
{\cal L}_{ENJL} \, &=& \,i\bar q  \hat D  q \,
+\,\frac{1}{4N_{c} \Lambda^2}\,\sum_{i,k=0,1} \left(a_{ik} 
\left[\bar q f_i q 
\bar q f_k q 
 - \bar q \gamma_{5}\tau^{a}
f_i q \bar q \gamma_{5}\tau^{a}
f_k q
 \right]\right.\nonumber\\
&&\left. - b_{ik} 
\left[\bar q \gamma_{\mu}\tau^{a} f_i q 
\bar q \gamma_{\mu}\tau^{a} f_k q 
 - \bar q \gamma_{5}\gamma_{\mu}\tau^{a}
f_i q \bar q \gamma_{5}\gamma_{\mu}\tau^{a}
f_k q
 \right]\right),
\label{MNJL}
\end{eqnarray}
where
\begin{equation}
f_{1}(\hat s) =2-3 \hat s \,; \qquad
f_{2}(\hat s) = -\sqrt{3} \hat s\,; \qquad \hat s \equiv
-\frac{\partial^{2}}{\Lambda^{2}}\,.
\end{equation}
With the CSB momentum cutoff $\Lambda$ one can specify the critical
values of coupling constants above which the dynamical CSB occurs,
 $a_{kl} \simeq 8\pi^2 \delta_{kl}$. Eventually, two multiplets of scalar,
pseudoscalar, vector and axial-vector mesons are generated in the vicinity of
these critical constants.

\section{Bosonization and technology}
First we consider the ECQM.
To find the characteristics of composite boson states in colorless quark
channels we incorporate auxiliary fields $\Phi$ in the scalar and pseudoscalar
channels, $\Sigma, 
 \Pi^a $, and in the vector and
axial-vector channel,
$W^{(\pm)a}_\mu $, 
and replace the
four-fermion operators by
\begin{eqnarray}
&&{\cal L}_{4-quark} = \bar{Q}\Gamma \Phi  Q
+ N_{c} \Lambda^2\frac{\Phi^2}{G_C}; \nonumber\\
&&\Gamma = 1; i\gamma_5 \tau^a; i\gamma_\mu\tau^a; i\gamma_5 \gamma_\mu\tau^a;
\quad C = S0, P1, V1, A1;
\end{eqnarray}
with an integration over new variables.

The actual value of
constituent mass $<\Sigma> = \Sigma_0$  in the ECQM is controlled \cite{11}
by the mass-gap equation (the minimum of the scalar effective potential)
\begin{equation}
\frac{ \Lambda^2}{G_{S0}}\left(\Sigma_0 - M_0 \right)
= - \frac{\Sigma^3_0}{4\pi^2} \ln\frac{\Lambda^2}{\Sigma_0^2}
\equiv \Sigma^3_0 I_0. \label{msg2}
\end{equation} 
Therefrom it is evident that the natural scale for the four-fermion
interaction is given by $\Sigma_0$ rather than by $\Lambda$ and it is
useful to redefine the related coupling constants:
$\bar G_C = G_{C}I_0 \frac{\Sigma_0^2}{\Lambda^2}$, characterizing
the weak coupling regime when
$\bar G_C \ll 1$.\\

The evaluation of physical characteristics in the 
ENJL model is similarly based on the bosonization in terms of auxiliary scalar
and pseudoscalar fields:
\begin{eqnarray}
&&{\cal L}_{4-quark} = \sum_{k=1}^{2} i \bar q\left(\sigma_k + 
i \gamma_5\pi_k^
+ \gamma_\mu\rho_{k,\mu} + \gamma_5\gamma_\mu a_{k,\mu} \right)  f_k(\hat s) q
\nonumber\\
&&+  N_{c}\Lambda^{2}\sum_{k,l=1}^{2}
\left(\sigma_k a_{kl}^{-1}\sigma_l + \pi_k^a a_{kl}^{-1}\pi_l^a
\rho_{l,\mu}^a b_{kl}^{-1} \rho_{l,\mu}^a + a_{l,\mu}^a b_{kl}^{-1} a_{l,\mu}^a
\right)
.
\label{bosonization}
\end{eqnarray}
Let us parametrize the matrix of coupling constants in a close vicinity
of tricritical point: 
\begin{equation}
 8\pi^2a_{kl}^{-1} = \delta_{kl} - \frac{\Delta_{kl}}{\Lambda^2},\qquad
|\Delta_{kl}| \ll \Lambda^2.
\label{couplQQM}
\end{equation}
The last inequality is equivalent to require
the dynamical mass to be essentially less than 
the cutoff.
After integrating out the quark fields one comes to the
bosonic effective action ${\cal W}(\sigma_k, \pi_k^a,\rho_{l,\mu}, a_{l,\mu})$.
The conditions on extremum of the effective potential, are given by
the mass-gap equations \cite{7}.\\

In both models the auxiliary
fields obtain kinetic terms from the quark loops and 
propagate, i.e. interpolate scalar
pseudoscalar, vector and axial-vector resonance states. 

The approximation is consistent for
a finite number of resonances if one retains only the part
of the soft-momentum expansion of quark loop together with a finite
number of vertices in the Quark Model effective action. 

This approach 
respects the confinement requirement, because one coherently 
neglects both the threshold part of quark loop  and
(the infinite number of)  heavier resonance poles. This is supported by
 the large-$N_c$ approximation which associates all momentum dependence in the
bosonized action  solely with meson resonances. 

An additional 
simplification is achieved with the help of the leading-log approximation when
one retains only the leading logarithm of the cutoff $\Lambda$ which is
a RG universal part of quark loops. In this approximation we lose somewhat in
precision
but gain the predictability and model independence. 

\section{Matching to QCD}
Let us exploit the constraints based
on chiral symmetry restoration at QCD at high energies.
We focus on two-point correlators of colorless 
quark currents
\begin{equation}
\Pi_C (p^2) = \int d^4x \exp(ipx)
\langle T\left(\bar q\Gamma q (x) \bar q \Gamma q
(0)\right)\rangle, 
\end{equation}
where $C = S, P, V, A$ and respectively $\Gamma = 1, i\gamma_5, \gamma_\mu 
\gamma_5 \gamma_\mu$.
In the chiral limit the scalar correlator  and the pseudoscalar one, as well
as the vector correlator and the axial-vector one, 
  coincide at all orders
in perturbation theory and also at leading order in the non-perturbative
O.P.E.\cite{11a}.
As the differences $\Pi_S - \Pi_P$ and $\Pi_V - \Pi_A$ decrease 
rapidly with growing momenta, one can
expect that the lowest lying resonances included into a Quark Model  
essentially saturate  the constraints from
CSB restoration.

For the ECQM and the ENJLM, in 
the scalar channel, one obtains the following sum rules \cite{11,14,14b},
\begin{eqnarray}
&&c_8+\frac{N_c\Sigma_0^2 I_0}{8\bar G_{S}}
-\frac{4\epsilon^2 N_c \Sigma_0^2 I_0}{8\bar G_{P}}=0,\label{srul1}\\
&&\sum Z_\sigma = Z_\pi +\sum Z_\Pi, \qquad  
Z_\pi = \frac{4<\bar q q>^2}{F_0^2},
\label{srul2}\\
&&\sum Z_\sigma  m^2_\sigma - \sum Z_\Pi  m^2_\Pi \simeq  
24 \pi\alpha_s <\bar q q>^2, \label{srul3}
\end{eqnarray}
where where the first one is applicable only to the ECQM.
It eliminates the contact term and fixes the bare constant $c_8$.
$ Z_\sigma, Z_\pi,  Z_\Pi$ stand for the residues
in resonance pole contributions in the scalar and pseudoscalar correlators.
In the minimal ECQM one reveals only one scalar meson with $ m_\sigma,
Z_\sigma$ and in the minimal ENJLM one has two scalar mesons $\sigma$
and $\sigma'$, with $  m_\sigma, m_{\sigma'},
Z_\sigma, 
Z_{\sigma'}$. In both models  there is a heavy pseudoscalar meson 
$\Pi \equiv \pi'$.
The last relation is essentially 
saturated by heavy pion and heavy scalar parameters.

In the vector channel one derives the relations \cite{14,14c}: 
\begin{eqnarray}
&&c_{10} = 0, \label{vsr1}\\
&&\sum f_V^2 m_V^2 =\sum f_A^2 m_A^2 + F_0^2,\label{vsr2}\\
&&\sum f_V^2 m_V^4 =\sum f_A^2 m_A^4,  \label{vsr3}\\
&&\sum f_V^2 m_V^6 - \sum f_A^2 m_A^6 \simeq  - 8 \pi\alpha_s <\bar q q>^2,
 \label{vsr4}
\end{eqnarray}
where the first one is applicable only to the ECQM to fix  the `bare' constant
$c_{10}$.

\section{Fitting and comparison of ECQM and\\ \mbox{ENJLM}}

 Let us  specify the input parameters.
We take
$F_0 = 90\ {\rm MeV}$,  $m_\pi^2 = 140\ {\rm MeV}$. We adopt \cite{13}
$\hat m_q (1\ {\rm GeV}) \simeq 6 \ {\rm MeV}$,
$<\bar q q> \simeq - (235 MeV)^3$,
and use the phenomenological value for the heavy pion mass
$m_\pi' \simeq 1300$ MeV 
\cite{10}. We also take the vector and axial-vector meson
masses,
$m_\rho = 770\ {\rm MeV}$ and  $m_{a_1} \simeq  1.2$
MeV, as known parameters. 

In the ECQM the fit of the scalar CSR rules 
(\ref{srul1})--(\ref{srul3})shows \cite{11,14} 
that they can be satisfied in the 
weak coupling regime of QCD as the 4-quark condensate correction is 
nearly irrelevant, being amount to 2\% of main contribution from the 
heavy pion.
We perform an optimal fit  applying in the vector channel only CSR
(\ref{vsr1}) and (\ref{vsr2}). For $m_\sigma
\simeq 1$ GeV one finds the chiral constant $L_8 \simeq 0.8\times 10^{-3}$. 
For $g_A =0.55$ one obtains the chiral constant
$L_5 =  1.2 \times 10^{-3}$ ($L_5, L_8$ to be compared with \cite{13}) 
and $\Sigma_0 \simeq 200$ MeV. Thus a rather heavy scalar meson is
provided by ECQM.

 Therefrom 
one derives that $\bar G_V \simeq 0.25,\, \bar G_A \simeq 0.2,$
$\bar G_S \simeq 0.11$, $\bar G_P \simeq 0.13$,
$\tilde g_A \simeq 0.66$, the bare pion coupling $f_0
\simeq 62$ MeV 
 and either
$\epsilon  \simeq 0.05$ or $\epsilon  \simeq - 0.51$. We see that 
the four fermion coupling constants $\bar G_S$ and $\bar G_P$ as well as
 $\bar G_V$ and $\bar G_A$ are slightly different and their values are 
 $\ll 1$,
signifying the weak coupling regime. 

However for the values $m_\rho = 770\ {\rm MeV}$, $m_{a_1} \simeq  1.2$
MeV and
$g_A =0.55$ the two vector sum rules (\ref{vsr3}),(\ref{vsr4}) 
are not satisfied.
They can be saturated only with additional vector multiplets.\\   

In turn, in the ENJL one finds that 
the leading
asymptotics (\ref{srul2}) represents the generalized $\sigma$-model relation 
and is automatically fulfilled.

As to the second constraint the possibility to satisfy it depends on the 
value of the QCD coupling constants $\alpha_s$. Eventually it
can be written \cite{14a} in terms of v.e.v.'s of two scalar fields, 
$\sigma_{1}, \sigma_{2}$:
\begin{equation}
m_{\sigma'}^2 -  m_{\pi'}^2 \simeq
2 \sigma^{2}_{1}
+\frac{4 \sqrt{3}}{3}\sigma_{1}\sigma_{2}
+ 6 \sigma^{2}_{2} \simeq \frac{6N_c\alpha_s }{8\pi} 
(\sigma_{1} - \sqrt{3} \sigma_{2})^2 .
\end{equation} 
As the left part is always positive there exists a lower bound for 
$\alpha_s \geq \frac{8\pi}{9 N_c}$  providing solutions 
of the constraint. The lowest value of $\alpha_s \simeq 0.9 $
is given by $\sigma_{1} = - \sqrt{3} \sigma_{2}$ and for these v.e.v.'s
one obtains the following splitting between the $\sigma'$- and 
$\pi'$-meson masses:
\begin{equation}
m_{\sigma'}^2 -  m_{\pi'}^2 \simeq 
\frac83 \sigma^{2}_{1} = \frac16 m^2_{\sigma};
\end{equation}
{\it i.e.} for $m_{\sigma} = 500 \div 600 MeV$ 
these masses practically coincide,
$m_{\sigma'} \simeq  m_{\pi'} = 1300 MeV$ and such a $\sigma'$-meson
may be identified \cite{10} with $f_0 (1300)$. 
However the above value
of $\alpha_s$ lies in the region of rather strong coupling where 
next-to-leading corrections to the anomalous dimension
of four-quark operator 
are not negligible, 
$\sim \frac{\alpha_s}{\pi} \sim 0.3$, and 
should be systematically taken into account to obtain 
a reasonable precision.

In the vector channel the mass spectrum exhibits the fast restoration of 
chiral symmetry. Namely with the above inputs for $m_\rho$ and $m_{a_1}$
 one obtains the splitting,
\begin{equation}
m_{a_1'}^2 - m_{\rho'}^2\simeq 
= \frac32( m^2_{\sigma'} - m^2_{\pi'}).
\end{equation}
Therefrom one predicts the mass of the excited $a_1$ meson to be
$m_{a_1'} \simeq 1500$ MeV, which has not been yet discovered.\\

Thus we have shown that both the ECQM and the ENJLM 
truncating 
low-energy QCD effective action can serve to describe the
physics of heavy meson resonances. The matching to nonperturbative QCD
based on the chiral symmetry restoration at high energies guarantees
the predictability of both models.\\ 

In the ECQM an optimal fit
supports the existence of rather heavy scalar quarkonium with the mass
of order 1 Gev. Meanwhile the ENJLM  inevitably
contains a rather light scalar meson with the mass $500-600$ MeV 
which is however not completely 
excluded by
the particle phenomenology \cite{10}, though this light meson is 
most probably a 2-pion or 4-quark bound state \cite{15}.\\

The fast convergence in the ENJLM of mass spectra and other characteristics
of heavy parity doublers entail their decoupling from the 
low-energy pion physics. For instance, let us calculate the dim-4 
chiral coupling constant \cite{16},
\begin{equation}
L_8 = \frac{F^4_\pi}{64 < \bar q q >^2} \left(\frac{Z_\sigma}{m^2_\sigma}
+ \frac{Z_{\sigma'}}{m^2_{\sigma'}}
- \frac{Z_{\pi'}}{m^2_{\pi'}}\right)
 \simeq \frac{F^2_\pi}{16 m^2_\sigma},
\end{equation}
when the CSR (\ref{srul2}), (\ref{srul3}) are imposed. The deviation
due to $\sigma'$- and $\pi'$-mesons \cite{14a} is less than 2\%. 
Thus this constant is essentially
determined in the ENJLM by the lightest scalar meson. Its phenomenological 
value, $L_8 = (0.9 \pm 0.4)\times 10^{-3}$ from 
\cite{13}  
accepts $m_{\sigma} \simeq 600 MeV$.

On the contrary, the same chiral constant in the ECQM contains a substantial
contribution (up to 50\%) from the heavy pion $\pi'$:
\begin{equation}
L_8  \simeq
\frac{F_0^2}{16}\left(\frac{1}{m^2_\sigma} + \frac{1}{m^2_{\pi'}}\right),
\end{equation}
with a good precision \cite{11}. 

Thus the two approaches encode quite a different dynamics of heavy resonances
responsible for the formation of structural chiral constants.\\

Some disadvantage of the ENJLM
as well as of the original NJL model
is that they assume large, critical values of  
four-quark coupling constants which
is difficult to justify with perturbative calculations in QCD. Respectively
the CSR matching has to be performed at a scale where the QCD 
coupling constant is rather large and the perturbation theory is unreliable.
The ECQM  seem to be free of these shortcomings.
However the final choice between them may be done by the
fit of a larger variety of meson characteristics which is in progress.

\vspace{0.5cm}

This work is supported by EU Network EURODAPHNE,  CICYT grant
AEN98-0431, CIRIT grant 2000SRG 00026 and 
by the Generalitat de Ca\-ta\-lu\-nya (Program PIV 1999). A.A. and V.A.
are supported  by   grants RFFI 98-02-18137 
and GRACENAS 6-19-97.

\end{document}